\documentclass[acmlarge]{acmart}
\makeatletter
\newcommand{\confshort}{\acmConference@shortname}
\newcommand{\conffull}{\acmConference@name}
\newcommand{\confdate}{\acmConference@date}
\newcommand{\confloc}{\acmConference@venue}
\AtBeginDocument{
  \fancypagestyle{firstpagestyle}{
    \fancyhead{}%
    \fancyfoot[C]{}%
  }
  \fancyhf{}
  \fancyhead[LO]{\@headfootfont\shorttitle}%
  \fancyhead[RE]{\@headfootfont\@shortauthors}%
  \fancyhead[LE]{\@headfootfont\footnotesize \confshort, \confdate, \confloc}%
  \fancyhead[RO]{\@headfootfont\footnotesize \confshort, \confdate, \confloc}%
  \fancyfoot[C]{}%
}
\makeatother
\acmBooktitle{\conffull\@ (\confshort), \confdate, \confloc}
\AtBeginDocument{%
  }

\copyrightyear{2026}
\acmYear{2026}
\setcopyright{cc}
\setcctype{by}
\acmConference[FAccT '26]{The 2026 ACM Conference on Fairness, Accountability, and Transparency}{June 25--28, 2026}{Montreal, QC, Canada}
\acmBooktitle{The 2026 ACM Conference on Fairness, Accountability, and Transparency (FAccT '26), June 25--28, 2026, Montreal, QC, Canada}
\acmDOI{10.1145/3805689.3812387}
\acmISBN{979-8-4007-2596-8/2026/06}

\usepackage{algorithm}
\usepackage{algorithmic}
\usepackage{csquotes}
\usepackage{multirow}
\usepackage{longtable}
\usepackage[table]{xcolor}
\usepackage{array}
\usepackage{colortbl}

\begin{document}

\title[The Hardness of Achieving Impact in AI for Social Impact Research]{The Hardness of Achieving Impact in AI for Social Impact Research: A Ground-Level View of Challenges \& Opportunities}

\author{Aditya Majumdar}
\email{adity@psu.edu}
\orcid{0009-0002-9413-2989}
\affiliation{%
  \institution{The Pennsylvania State University}
  \city{University Park}
  \state{Pennsylvania}
  \country{USA}
}

\author{Wenbo Zhang}
\affiliation{%
  \institution{The Pennsylvania State University}
  \city{University Park}
  \country{USA}}
\email{wjz5120@psu.edu}
\orcid{0009-0002-4589-6093}

\author{Kashvi Prawal}
\affiliation{%
  \institution{The Pennsylvania State University}
  \city{University Park}
  \country{USA}}
\email{kjp60060@psu.edu}
\orcid{0009-0005-9508-831X}

\author{Amulya Yadav}
\affiliation{%
  \institution{The Pennsylvania State University}
  \city{University Park}
  \country{USA}}
\email{amulya@psu.edu}
\orcid{0009-0005-4638-9140}

\renewcommand{\shortauthors}{Majumdar et al.}

\begin{abstract}
  AI for Social Impact (AI4SI) is an emergent field harnessing interdisciplinarities between the fields of artificial intelligence (AI), machine learning (ML), and the social sciences to address societal issues aligned with the United Nations Sustainable Development Goals (UN SDGs), such as universal healthcare, climate action, etc. Despite AI4SI’s rising popularity, achieving tangible, on-the-ground impact remains a significant challenge. In particular, identifying collaborators open to co-designing and deploying AI4SI-based solutions in real-world settings is often difficult. Thus, many projects stall at the proof-of-concept stage, unable to scale to production-level deployment. Drawing on twenty-six AI4SI researchers’ interviews, primarily from academic institutions though also including some industry researchers and practitioners, and the authors’ own lived experiences, this paper employs thematic analysis to highlight structural, organizational, communication, collaboration, and operational challenges hindering socially impactful AI4SI deployments. While there are no easy fixes, the authors synthesize best practices and actionable strategies from interviews and personal experiences, positioning this paper as a practical guide for AI4SI researchers and organizations pursuing socially impactful collaborations\footnote{We note that our findings are most directly applicable to academic research groups in the global north, as governmental, startup, and global south researchers' perspectives are underrepresented in our sample.}.
\end{abstract}

\ccsdesc[500]{Applied computing~Law, social and behavioral sciences}
\ccsdesc[500]{Computing methodologies~Machine learning}
\ccsdesc[500]{Computing methodologies~Artificial intelligence}

\keywords{Participatory AI, AI for Social Impact, AI for Social Good, Semi-structured Interviews, Thematic Analysis}

\received{13 January 2026}
\received[Revised]{25 March 2026}
\received[Accepted]{16 April 2026}

\maketitle

\section{Introduction}

The number of Artificial Intelligence for Social Impact (AI4SI)\footnote{AI4SI is also commonly referred to as AI for Social Good (AI4SG). Though these terms are used interchangeably, we will consistently use AI4SI to refer to this research field in the rest of this paper.} projects has risen rapidly in recent years. Between 2018 and 2023, the number of such projects tripled from 170 to over 600 projects \cite{Bankhwal_2024}.
Most projects in this field harness AI to tackle one of the 17 UN SDGs, 
which include equity in healthcare \cite{saheli2023,tabar2020identifying,liang2020let}, public safety \cite{armor2008}, education \cite{education2021}, social justice \cite{vajiac2024preventingevictioncausedhomelessnessmlinformed}, and other areas of global concern. Motivated by a philosophy of ``\emph{helping the helpers}'', these AI4SI projects are often carried out by AI researchers in close collaboration with 
partner organizations (POs) such as non-profits, hospitals, government agencies, and other community stakeholders \cite{Matias_Hassidim_Nelson_2025}. Thus, they require deep collaboration across disciplinary, organizational, and institutional boundaries. Unlike traditional AI and ML research, these projects are rooted in complex social ecosystems, and their results must align with the practical constraints, workflows, and values of the PO, and the low-resource community they serve. 
Therefore, expectations for an AI4SI project to be considered a “success” are stringent. Such projects must go beyond theoretical or empirical performance gains to demonstrably drive meaningful social impact within the low-resource communities they aim to serve.

Unfortunately, achieving ``success” in the AI4SI space often proves challenging for a variety of reasons. Despite growing interest, 
researchers often struggle to move AI4SI solutions beyond proof-of-concept or short-term pilots to long-term, at-scale deployment for their POs.
A recent study found that 72\% of AI4SI projects never leave the research and development phase \cite{Bankhwal_2024}. Although technical issues can contribute to this challenge \cite{zou2020concordance}, they are rarely the sole reason AI4SI solutions fail to transition into real-world deployment. Recognizing this, several frameworks \cite{Kshirsagar2021BecomingGood,Tomasev2020AIforSocialGood,Floridi2020HowToDesignAI} have been proposed to help researchers design 
responsible AI4SI technologies and sustain collaborations (see section \ref{sec:related_works}). Although these frameworks offer insightful guidance, a key limitation in the current literature is that they abstract away from the lived realities of field deployment, making it nebulous how such 
frameworks translate to interdisciplinary, resource-constrained environments.  
Recent work by \citet{hongjin_lin} highlights this challenge from the perspective of POs, showing that while community partners may expect tangible deployment outcomes, only a small fraction of AI4SI projects reach that stage. In contrast, our study approaches this problem from the AI4SI researchers’ perspective, and asks: 
\emph{What factors prevent AI for Social Impact projects from being deployed in real-world settings?}

To answer this question, we conduct a series of semi-structured interviews with active experienced researchers with successful collaborations in the field of AI4SI, the majority of whom are based at academic institutions in the global north, though our sample also includes industry researchers and one startup practitioner. Our findings are therefore most directly applicable to the academic research group context. Through these interviews, supplemented by the authors’ own AI4SI research experiences, we identify structural, organizational, communication, collaboration, and operational hurdles that hinder deployment. 
As a result, this paper describes the challenges encountered in developing and deploying AI4SI applications for PO-based stakeholders which serve low-resource and underserved communities with limited financial and human resources. 
While there are no easy fixes to the challenges described in this paper, we outline best practices from interviews and the authors' experiences that may help reduce their impact on AI4SI collaboration outcomes.
In turn, we hope this paper serves as a helpful reference for both new and experienced AI4SI and Responsible AI practitioners, as well as POs seeking to integrate AI into their operations. 

\section{Related Work}
\label{sec:related_works}

\textbf{Conceptual Foundations \& Design Principles for AI4SG}: In their 2020 chapter, \citet{Floridi2020HowToDesignAI} look across twenty-seven AI‐for‐Good projects, from disaster‐response chatbots to health‐monitoring tools, and extract seven guiding principles.  They emphasize starting with a problem that truly matters, bringing stakeholders into the design from day one, and building in ethical checks at every turn. Each principle is grounded in concrete practices, such as community co‐design workshops and automated fairness audits, to help teams translate ideals into real-world impact. Around the same time, \citet{Tomasev2020AIforSocialGood} surveyed efforts aligned with the UN’s 17 SDGs.  By examining case studies in agriculture, public health, and beyond, they distilled ten practical recommendations, everything from setting clear success metrics to embedding an ethics review board.  Their work highlights how deep, ongoing collaboration with domain experts keeps projects on track and avoids the “\emph{tech for tech’s sake}” trap. Building on these ideas, \citet{Kshirsagar2021BecomingGood} take a more hands‐on approach.  Instead of broad principles, they break AI4SG collaborations into four daily workflows, namely, communication, data management, modeling, and impact assessment, and draw out eleven lessons.  
Two case studies showcase how teams build lasting, trust-based partnerships. Unfortunately, while these studies offer useful frameworks, they often present ideal workflows that obscure how things unfold in messy real-world settings.

\noindent \textbf{Deployment Challenges \& Pipeline Planning}: 
Even strong AI4SI ideas can falter before reaching deployment due to structural and organizational constraints. Complementary to our work, \citet{hongjin_lin} explores the challenges faced by POs in AI4SI collaborations. They highlight that although POs enter AI4SI projects expecting deployed systems, only a fraction of those projects reach that stage due to the various hurdles such as funding structures and power asymmetries rather than just technical limitations. Their findings provide important PO-centered evidence of deployment challenges in AI4SI collaborations.
Next, \citet{Saxena_2025} propose a framework to identify ``\emph{AI Mismatches}'', which occur when model performance falls short of real-world task requirements and leads to harm. Based on an analysis of 774 AI cases, they develop seven matrices that help teams evaluate feasibility, risk, and equity early in development. They argue that many failures arise from flawed problem formulation rather than technical mistakes, and that early design stages are the best time to surface ethical concerns. Case studies show how the framework helps prevent harmful AI deployments.
Finally, \citet{Perrault2021ArtificialIntelligence} propose a four‐stage “\emph{data‐to‐deployment}” pipeline for AI4SI: (1) problem identification and immersion (engaging stakeholders to define real-world context); (2) prediction (developing accurate models); (3) prescription (translating models into actionable decisions); (4) field testing and deployment (evaluating solutions through real-world pilots and feedback). Examples in flood prediction and maternal health show how this approach keeps prototypes from failing in practice.
In our paper, we build on top of these studies by providing grounded insights from researchers navigating day-to-day challenges of deployment in AI4SI research. \\ 
\noindent \textbf{Similar Works in Allied Fields}: AI4SI research is part of a long lineage of work examining how computing technologies address societal and developmental challenges. The Information and Communication Technologies for Development (ICT4D) literature has long documented challenges in translating technological innovation into sustained societal impact. Prior work identifies recurring barriers such as misalignment between system design and real-world contexts, technology-centric approaches that overlook social and institutional realities, and limited community engagement during system development \cite{Heeks2002,Heeks2008}. \citet{Walsham} traces the history of ICT4D and notes that institutional incentives often discourage the practitioner engagement needed for real-world impact. \citet{Harris2016} similarly finds that although ICT4D researchers value impact, institutional structures disincentivize user engagement and broader dissemination.\\
Similar critiques were raised in the broader Tech for Social Good (T4SG) and Data Science for Social Good (DSSG) space. For example, \citet{pal2017fallacy} argued that many “\emph{design for good}” initiatives risk treating marginalized populations as objects of technological experimentation rather than engaging deeply with the social contexts in which technologies are deployed. Recently, \citet{aula2023stepping} highlighted how many challenges in contemporary tech-driven initiatives reflect the long-standing tensions between technological innovation, institutional constraints, and real-world deployment.\\
Building on prior work highlighting challenges in technology-for-good collaborations, our study contributes to this line of work by empirically documenting how these challenges persist in AI-driven collaborations. Furthermore, we also highlight how these challenges may be intensified by AI-specific dynamics such as the rapid pace of AI/ML research.
\section{Methods}
\label{sec:methods}
To answer our research question, we conducted semi-structured interviews with experienced AI4SI researchers who have worked or are currently working in the field. Research communities have varying definitions of the phrase AI4SI, hence we use the following normative definition of AI4SI in this paper: \emph{AI for Social Impact refers to the development and deployment of AI systems that are co-designed with impacted communities, prioritize equity and inclusion, and aim for measurable, long-term improvements in societal well-being for historically neglected social challenges, particularly within the low-resource, underserved, or marginalized contexts.} 
Thus, while some may argue that much AI/ML research leads to long-term social impact (e.g., ChatGPT), this paper focuses specifically on AI technologies developed to address challenges faced by low-resource communities.
We believe this focus allows us to center often-overlooked beneficiaries of AI technologies, including but not limited to homeless individuals \cite{homelessnessvajiac2024_example,yadav2015preventing,yadav2016using,yadav2016pomdps,yadav2017using, yadav2017maximizing, yadav2018bridging,rice2018piloting,yadav2021optimal,tabar2020identifying,wilder2021clinical,dou2021harnessing,rice2021peer}, subsistence low-literate farmers \cite{farmersingh2024_example, tabar2021plan, indebtfarmers,tabar2022mitigating}, malnourished children \cite{malnourished_example}, individuals at risk of eviction \cite{evictiontabar2022_example, Maryam2}, and expectant or new mothers in rural areas \cite{expectantmothers_example, 10.1609/aaai.v37i12.26709}.


\noindent \textbf{3.1 Recruitment and Data Collection}:
The study was conducted at the main campus of an R1 university in the Mid-Atlantic region of the United States. 
To recruit participants, we used a combination of purposive and snowball sampling. We first employed purposive sampling to identify ‘seed’ participants:  
(1) We initially selected all corresponding authors of papers published in the past five years at the AAAI Workshop on AI for Social Good, AAMAS Workshop on Autonomous Agents for Social Good, KDD Workshop on Data Science for Social Good, AAAI AI for Social Impact Track, IJCAI
AI for Society Track, EAAMO, and KDD Applied Data Science Track. 
(2) We then extended recruitment using a researcher-led (citation-network-informed) snowball sampling strategy by analyzing citations and co-authors of these ‘seed’ participants;
co-authors who had also published in the above AI4SI venues were added as prospective participants. (3) All these prospective participants were invited to be a part of our study via email. 

Prospective participants' emails were sourced from the directories of their host institutions. The recruitment email included a hyperlink to a published AI4SI paper authored by the recipient and a hyperlink to schedule an interview time based on the prospective participant's availability. Up to three weekly reminder emails were sent, and participants who responded were selected for the study. Through this process, we were able to enroll 
$N = 26$ participants, with interviews lasting  $\sim$ 43 minutes on average.
The data collection began in March 2025 and ended in August 2025. 

The interviews were conducted and recorded via Zoom. Before the interview, participants were informed of the study’s motivation and interview overview, then asked to provide verbal consent for participation and recording.
The interview transcripts were transcribed verbatim and anonymized for analysis. Participants were informed that participation was voluntary and that they could withdraw at any time without penalty. 
Initially, participants were asked to elaborate on the AI4SI project referenced in their recruitment email and to describe the current state of their technology.
They were then asked to share experiences from other AI4SI projects, especially those that failed to deploy or had unexpected outcomes.
The interview concluded with a request for suggestions to improve collaboration models in the AI4SI space. Afterwards, participants completed a demographics survey. 
Our interview protocol was approved by the Institutional Review Board and conducted in accordance with the ethical guidelines of the host institution. Aggregate participant demographics are reported in Table \ref{tab:demographic} in Appendix \ref{sec:appendix_demographics}.

\noindent \textbf{3.2 Data Analysis}:
We adopted a thematic analysis approach to analyze the interview transcripts \cite{Auerbach_Silverstein_2003, Braun2019}. Initially, the first author did multiple passes of the transcribed text to understand the breadth of topics covered by the participants. Following this, we used collaborative qualitative analysis \cite{Richards_2017}. The first and second authors independently analyzed $\sim$1121 minutes of audio recordings by performing open coding on the transcripts. Using an inductive approach, they let the themes emerge from the text. A total of thirty-three preliminary codes were generated in this phase. The two authors then met to compare and refine their codes, discussed interpretations, and resolved any differences. Building on this, we moved to the second phase of analysis, using theoretical coding \cite{Hennink_Hutter_Bailey_2019} to organize the initial codes into higher-order themes, presented in Section \ref{sec:challenges_uncovered}.
This process was iterative, involving ongoing refinement of codes and themes over multiple weeks to establish validity and reduce bias in our coding process. Any disagreements were resolved through rigorous discussions until a consensus was reached among the authors. Throughout this process, we maintained an audit trail documenting coding decisions and memos to ensure the validity of the generated codes \cite{Creswell_2000}. All of the generated codes were analyzed using Taguette \cite{Rampin2021}, a qualitative data analysis software. In the findings, quotes from the $i^{th}$ participant (and author) are labeled as \(P_i\) (and  \(A_i\)), respectively. 
Participants discussed 33 AI4SI projects, in addition to the authors’ lived experiences from 5 projects, for a total of 38 projects summarized in Table \ref{tab:AI4SI} in Appendix \ref{sec:appendix_projects}.
\section{Challenges Uncovered}
\label{sec:challenges_uncovered}

We organize our findings around four distinct challenge themes identified through our analysis, derived from both participant accounts and the authors’ lived experiences. 


\subsection{Structural and Organizational Challenges} 
Through our analysis, we identified four types of structural and organizational challenges:

\noindent\textit{4.1.1}$\>$\textit{Misaligned Incentives in Academia.} \label{sec:4.1.1}\\
\noindent \textit{A) Pervasive “Publish or Perish” culture: }Due to a pervasive “publish or perish” culture in academia, researchers interested in AI4SI projects often find themselves at conflicting crossroads: the \emph{desire}
to produce meaningful, context-aware, deployable solutions with measurable real-world impact v/s the \emph{need} to publish a high volume of papers
at well-reputed, competitive venues. To achieve ``success”, AI4SI projects often require long-term partnerships with community stakeholders, extensive engineering, and field work to design systems with real-world relevance and impact. These realities of field-driven AI4SI work, which rarely lead to high-volume publications, often misalign with academic success metrics.
As a result, the lack of appropriate incentive structures in academia often leads AI4SI projects to be deprioritized in favor of “lower hanging fruits” or more ``easily publishable'' work.
$P_{1}$ candidly reflected on this tension: 
\begin{displayquote}
\textit{I think that one problem [...] is the culture, [...] the incentives of CS academia have become really high number of publications focused. [...] That's a very bad thing for the field and for the depth of work, and for the ability to do the kind of work that I would like to do, and would like to see people doing in the [AI4SI] space. [...] We really need to try to figure [this] out, because the incentives are bad in some ways.}
-- $P_1$
\end{displayquote}

Other participants also noted how the time-intensive nature of these collaborations clashed with the rapid, competitive pace of academic publishing. $P_5$ noted how this pressure manifests for PhD students working on AI4SI projects: 


\begin{displayquote}
    \textit{As a PhD student in AI/ML, you probably know how fast the publication cycle is [...], [while your peers] can publish like 3 papers a year [without worrying about AI4SI field work]; but if you work on this AI4SI project [...] you may be forced into doing engineering work for months trying to get [your AI4SI solution] deployed, with no clear roadmap to getting a paper. This is a true blocking factor for a PhD student. -- $P_5$}
\end{displayquote}



\noindent \textit{B) Conventional Expectations of Novelty: }$P_4$ further illustrated how conventional expectations of novelty in AI academic research shaped their early-career publishing strategy, where they were forced to mask their AI4SI field work as part of an ``applications'' section, to increase the chances of their work being accepted. 
\begin{displayquote}
    \textit{Since there was no such venue to publish [AI4SI] papers, the only way to do this was to smuggle it in [a conventional AI venue] by writing a methods paper and then putting in [AI4SI work] as an application. But that wouldn't occupy as much importance in terms of review. And so, a lot more effort [was spent] trying to add unnecessary methodological improvements, to things that really were not that useful.} -- $P_4$
\end{displayquote}


Author $A_4$ shared similar experiences, noting that the need to accommodate real-world constraints in AI4SI often produced complex problem formalizations that limited theoretical insight. To satisfy expectations of traditionally minded reviewers, $A_4$ included trivial hardness theorems to align with academic norms rather than advance AI4SI goals.

In addition, many AI4SI projects require high levels of engineering work to reach deployment, which rarely falls within conventional definitions of novel research, disincentivising researchers from pursuing such projects (or mentioning this effort in their papers). $P_5$ stated:\textit{``PhD students are here to solve research problems. And the engineering work is additional work that they need to do beyond their research work. It's time-consuming"}.

\noindent \textit{C) A Careful Balancing Act: }Navigating this tension requires researchers to balance a difficult trade-off between technical novelty and ensuring real-world utility and impact. $P_6$ reflected that focusing solely on novelty (cared for by the broader research community) can undermine the practical utility of the system:

\begin{displayquote}
    \textit{I feel that as academics we need to publish and let people know what we have done. So it's a trade-off and a tricky line to figure out a novelty, but also make sure it gets deployed, and it's useful 
    [...] because if I had just worried about novelty, then [PO] may not have even used our methods.} -- $P_6$
\end{displayquote}



Due to these structural disincentives, academics are less likely to pursue/maintain projects that do not align with conventional expectations in the broader CS community. Even 
though an AI4SI project may have the potential to mitigate a pressing social issue, it is deprioritized since the extensive time, engineering, and deployment efforts aren't recognized at an institutional 
level. As a result, valuable work is set aside not due to lack of relevance, but because it does not fit into prevailing academic incentive structures. As $P_4$ emphasized, ``\emph{incentive structures not just influence individual motivation, but also shape which projects are considered viable, making projects difficult to justify when their labour/outcomes fall outside dominant academic incentive structures, despite their societal importance.}" 

\noindent\textit{4.1.2} $\>$ \textit{Misaligned Incentives in Partner Organizations.} \\
\noindent \textit{A) Extra-Role Expectations: }
While misaligned academic incentives challenge AI4SI deployment, POs face their own barriers. Often operating with limited budgets and small teams, including volunteers, POs 
rarely have the capacity or formal mandate to engage in long-term AI4SI research collaborations. As $P_6$ mentioned, ``\emph{this is a new thing they have to do. Their current job does not have this; they aren’t getting paid for this. [...] This isn't part of their job description.}" Author $A_1$ echoed this sentiment, noting that reliance on unpaid non-profit volunteers across a 12-hour time difference slowed coordination, eventually making several of their AI4SI projects unsustainable.


\noindent \textit{B) Divergent Priorities: }In some cases, the lack of sustained engagement stems from misalignment between the research focus and the PO's actual priorities, a finding that supplements \citet{Saxena_2025}. As $P_{14}$ mentioned, \textit{``the problem that the technologists think that they're solving may not be the problem that the community has or wants to be solved"}. $P_5$ further added that during the initial phases of a collaboration, when early problem formulation efforts settle on a compromise that is technically solvable but not truly central to the PO's mission, they lack \textit{``the incentive to invest time, energy, and human resources''} required to carry the project through deployment. 

Even when a relevant problem is identified, misalignment over what constitutes a core need can persist across PO levels. While leadership may express enthusiasm for integrating AI, front-line workers may view such tools as irrelevant or even burdensome. $P_4$ shared their experience from a wildlife conservation experience:

\begin{displayquote}
    \textit{We went on a trip with [PO to discuss AI-enabled] wildlife conservation. And a high-level secretary said, ``sounds like an awesome idea, let's [do it]!" But then, when we go to the rangers on the ground, they said, ``we just need better shoes. We just need better guns. We don't need AI. [...] we have these old guns, and the poachers come with better guns". And they're just completely opposed to AI.} -- $P_4$
\end{displayquote}

 This perspective is mirrored by recent work on recruitment in participatory AI, which highlights that long-term success of collaboration depends not only on technical alignment, but also through early establishment of relationships, shared priorities, and expectations \cite{relAItionship_building}. Participants emphasized that without establishing a clear vision and strong early buy-in, these collaborations are unlikely to result in deployment. As $P_2$ noted \textit{``you sort of need that [clear vision] otherwise [...] at some point, things will falter"}. Hence, 
 alignment on priorities must be established early, like $P_6$ shared: 

 \begin{displayquote}
    \textit{In the first few parts of the project, it's very important to ask the hard questions, so people are clear about what it takes to be successful, and are they really serious about it? Because if there is no buy-in from the stakeholders, these projects cannot succeed.} -- $P_6$
 \end{displayquote}

In summary, deployment is not just a technical task, but is a process which requires deep mutual buy-in from the beginning, since without it, even well-intentioned AI4SI projects cannot move beyond the prototyping phase.

\noindent\textit{4.1.3} $\>$ \textit{Funding.} \\
\noindent \textit{A) Insufficient Funding for AI4SI: }
Even when researchers and POs align on a mutually beneficial problem, limited funding can still impede project progress. $P_5$ noted that \textit{``It's not that easy to get a lot of funding for this kind of work. [...] The nonprofit is also resource-constrained, so they expect some funding support to support their human efforts to work on this AI project''}, highlighting that both researchers and POs face financial constraints, with POs often relying on research teams to support project costs beyond initial prototyping. 
This challenge was echoed by $P_6$, who, reflecting on a project at the onset of COVID-19, noted that the lack of immediate funding required them and their students to volunteer time to support POs, an approach they said is unsustainable in the long term. $P_6$ recalled:  \textit{``obviously there was no money immediately.[...] So you had to do pro bono or volunteering work. But you need to quickly figure out that this is not sustainable long term. You need to have a funding stream."} $P_{12}$ further mentioned that the challenge was less about coming up with actionable ideas and more about the hidden costs, something the funding agencies fail to recognize:


\begin{displayquote}
  \textit{The real barrier is not the lack of very good ideas. It’s more the lack of understanding [from funding agencies] of how much it’s going to cost, not just dollars for developers, but project management, infrastructure, long-term maintenance, all that kind of stuff. That’s often where I see a lot of things fail.} -- $P_{12}$
\end{displayquote}



 $P_{15}$ highlighted the challenge of limited access to proper hardware when initiating collaborations, \textit{``[POs] also [have] limited financial resources, [...] limited compute resources. [...] So there's no way that they can build [or host] the AI''}.
Even when POs wish to continue, this constraint prevents meaningful long-term integration into their systems.
$P_2$ shared a scenario where they developed an intervention but the PO couldn't deploy it as they didn't have the necessary funds: 

\begin{displayquote}
    \textit{Even aside from the AI part [...] the object-level intervention itself was being provided by the research team. If we could have gone to them and said, like, ``Hey, we have the capacity to continue delivering this intervention", I think they would have loved for us to do that, and they would have worked with us to bring that into their space. But it's not something that they could have provided themselves at that time.} -- $P_2$
\end{displayquote}

    
Surprisingly, even in the case where the research group can provide funding to maintain the intervention, they are rarely sufficient for long-term deployment. As $P_4$ reflected: 

\begin{displayquote}
    \textit{We gave grants to these wildlife conservation organizations to hire people to try to do this [maintenance work], but they couldn't. This just needs a level of support and organization that our students cannot [provide], [...] Students have to do their own research, and they can't be asked to do maintenance.} -- $P_4$
\end{displayquote}

Ultimately, without a long-term funding model, even high-potential project deployments can run out of momentum. As $P_6$ precisely stated \textit{``if funding is not figured out, it may not fail immediately, but it will fail eventually, because at the end of the day somebody has to pay for it."}

\noindent\textit{4.1.4} $\>$ \textit{Institutional Inertia and Disciplinary Gaps.} 

\noindent \textit{A) Red-Tape at Institutional Organizations: }
Another hurdle AI4SI projects often face is institutional processes misaligned with the pace of field-driven deployment, which can slow progress. $P_5$ noted that both researchers and POs can be under-prepared when navigating procedural red-tape at their host institutions: 

\begin{displayquote}
    \textit{There are a lot of steps that we need to go through, and this could scare both the nonprofit organization, and the researchers. For the nonprofit organizations, maybe they're not familiar with this, so they don't even know they would need to spend time on this and the same thing on our end for the students.} -- $P_5$
\end{displayquote}

This view was echoed by $P_2$, who mentioned administrative processes such as data use agreements often operate on timelines which cannot keep up with research teams' pace: ``\textit{Data use agreements can be a real pain [...] it all has to go through the lawyer's office, it can take forever. We just try to start them as soon as possible, and then keep asking the lawyers until they're done.}''


\noindent \textit{B) Improper Interdisciplinary Training: }
Beyond organizational delays, research teams often lack the interdisciplinary training needed for fieldwork, as most students are trained in CS or ML rather than econometrics, human-centered design, or software engineering. As $P_4$ noted: 

\begin{displayquote}
    \textit{People ask me, why don't people [who] do AI, [just] know how to do [AI4SI]? [...] because there's a lot of econometrics training and a lot of statistical analysis, and a lot of field experiments, and what social workers do, and all sorts of things that we need to do [or know] to interface AI with society.} -- $P_4$
\end{displayquote}

On a similar note, $P_3$ mentioned how people working in ML may not have abilities to develop software applications:
\textit{``Someone might need an Android app, someone might need an iPhone app[...]. People doing ML might not have the skills to build websites or apps."}
Together, institutional inertia and gaps in interdisciplinary training leave many AI4SI projects stalled in early stages, not due to a lack of will, but because surrounding structures fail to support implementation.

\subsection{Communication and Understanding Gaps}
\label{sec:4.2}

Spanning multiple disciplines (e.g. CS, medicine, public policy) and sectors (e.g. academia, healthcare), AI4SI projects often encounter communication breakdowns and misunderstandings that collaborators must learn to navigate.

\noindent \textit{A) Contested Meanings - AI perceptions and Disciplinary Jargon:} 
Media narratives around AI have fueled widespread misunderstandings, inflating expectations and concerns. 
As $P_1$ aptly put it:
``\textit{I think it's [...] unfortunate in some ways, because I think AI 
    [...] gets painted with a very broad brush, and the kind of AI that I'm talking about is often not the kind of AI that POs are worried about.}''
This misunderstanding further manifests in contradictory perceptions. $P_{1}$ noted:
\begin{displayquote}
    \textit{There's fairly deep skepticism about AI, while at the same time, there's also this sense of like, ``Oh, wow! Can it do great things for us?" [...] Educating people about what AI is and what are the kinds of things that we can do is a big thing that we need to do.} -- $P_1$
\end{displayquote}

Hence, without a grounded understanding of what these systems can do for the PO, initial conversations between collaborators can be quite tricky and time-consuming. As $P_2$ stated: ``\emph{It takes a while because they do not know anything about what AI could do, so they don't know what to ask for, and you don't know enough about their domain to tell them what they should be asking}".
In general, challenges in communicating across disciplinary boundaries can stall the progress of these projects. As $P_{5}$ mentioned 
\textit{``figuring out how to communicate across these interdisciplinary boundaries is super difficult"}, particularly when foundational concepts and jargon differ. This is when participants mentioned the importance of long-term sustained collaborations, for example, $P_1$ shared \textit{``that's why having relationships that go on for a long time are good, because then you come to a shared understanding of what's happening and what you can do."}

\noindent \textit{B) Importance of Internal ``Champions": }
One strategy participants employed was identifying PO collaborators with shared expertise. $P_8$ mentioned \textit{``it was very important that there be a champion on the side of the social impact organization to work with us, provide input, feedback and be in a position to operationalize some things"}. Furthermore, $P_5$ explained: 

\begin{displayquote}
    \textit{Our sense was that if we had somebody on the inside who was trained in economics or game theory, or some probabilities and things like that, then we found it very easy to talk to them, and then they would be the ones who champion the system internally.} -- $P_5$
\end{displayquote}



In summary, clear and consistent communication is critical for such collaborations, as without shared understanding, even well-intentioned projects can falter. As $P_6$ said, \textit{``it's important at every point of time that you keep communicating. Keep telling what you have in mind, understand their issues and constraints, otherwise it can quickly break down".}

\subsection{Collaboration Hurdles}
\label{sec:4.3}
Through our analysis, we identified two types of collaboration hurdles: forming strategic partnerships and trust issues.

\noindent\textit{4.3.1} $\>$ \textit{Strategic Partnerships.}

\noindent \textit{A) Buy-in from PO Leadership: }
Successful deployments in AI4SI rely on not only the technical feasibility, but also on fostering relationships with collaborators that have the authority and commitment to allow the integration of AI4SI-based interventions in the PO's operational pipeline. Participants mentioned the importance of having institutional buy-in from the PO's side and someone from the organization with actual decision-making power. $P_2$ explained: 

\begin{displayquote}
    \textit{And the sort of the deeper thing is like a lack of institutional capacity and buy-in to get to implementation. Because at the end of the day, deploying something requires time, commitment, and someone with sufficient decision-making power being on board right in the partner organization.} -- $P_2$
\end{displayquote}

In situations where the commitment from leadership was missing, the participants found themselves waiting for extended periods of time, and ultimately being forced to restart with another PO. $P_6$ reflected on their experience: 

\begin{displayquote}
    \textit{I remember one hospital we were working with, [...] we waited for one and a half years, and that guy was not even responding to our emails then, and we were stuck. And then we found out another right partner in another hospital and within 3 months we had applied for an NIH grant.} -- $P_6$
\end{displayquote}

\noindent \textit{B) Collaboration with Intermediaries: }
Some researchers addressed this challenge by working with “intermediaries,” often scholars from disciplines such as social work or medicine who have long-standing relationships with POs, helping expedite logistics and build trust. 
$P_2$ described their AI for HIV prevention work, noting that while they developed the algorithm, a separate social work research team implemented the interventions in drop-in centers:

\begin{displayquote}
    \textit{The social work research team had like a long-running partnership with the drop-in centers, and so they would be able to come to the center and recruit people for the study and use space at the center to offer the interventions themselves, [...] like the labor itself is all being supplied by the social work research team under their own sources of funding.} -- $P_2$
\end{displayquote}

Similarly, $P_1$ noted the value of working with an intermediary, who could directly engage decision makers in the PO:

\begin{displayquote}
    \textit{I'm a computer scientist, but my partner in crime [partner's name], his work's in public health, and he sits on a bunch of advisory boards and committees, and he brings the result of this research to the people who are making decisions right? And so he talks to them and [...] helps to inform decision makers.} -- $P_1$
\end{displayquote}

However, such collaborations introduce their own challenges, as interdisciplinary work requires agreement on shared incentives, including where and how to publish. $P_1$ acknowledged this by stating: 

\begin{displayquote}
    \textit{How does one negotiate like, where are we going to try to publish this? What are the kinds of things that we can do in terms of publication? [...] I'm grateful that [partner's name] is very happy with publishing a whole bunch of stuff in AI and CS venues. There are many other disciplines where that would not be okay, or where people would care about that. } -- $P_1$
\end{displayquote}

In addition, despite their advantages, such partnerships can become bottlenecks for new AI4SI researchers, who may not know who these intermediaries are or have access to their networks. They could therefore find themselves excluded from conversations that rely heavily on pre-existing trust. $P_{26}$ noted \textit{``I think this can also result in difficulty for new people, especially new people who don't already have connections to well-connected people''}. Overall, strategic collaborations require not just interdisciplinary understanding, but also shared priorities, mutual trust, and access to PO's key decision makers which enable integration of interventions in their workflows.

\noindent\textit{4.3.2} $\>$ \textit{Trust Issues.} \\
\noindent \textit{A) Skepticism Toward Researcher Motives: }
Another recurring theme was the difficulty of building trust with POs, particularly early in collaborations, when skepticism about researchers’ motives is common. $P_5$ pointed out \textit{``they [POs] don't trust you because they feel like you just come and get a paper, and then you're gone."} Previous negative experiences may also account for an organization's skepticism towards collaborating with researchers. $P_1$ explained that schools and local governments may be especially cautious due to reputational risks: 
\begin{displayquote}
    \textit{It is difficult to build trust, because they have repeatedly had experiences where they've worked with academia, and it hasn't gone well, or there's been some kind of information spillage [...] then they get bad press because something happens in your local school division -- it's a big story.} -- $P_1$
\end{displayquote}

To address these concerns, participants emphasized creating explainable models and communicating long-term commitment intent early in the collaboration.
$P_{23}$ mentioned \textit{``we deliver a data visualization, or some [interpretable] interface in the first 3 months [...] it actually helps the community partner, [and] keeps them interested in the project''}. 


\noindent \textit{B) Territorial Disputes: }
Author $A_4$ also encountered mistrust from practitioner-oriented stakeholders. In an AI4SI project on locust swarm forecasting, the work was shelved after a proof-of-concept was presented to a PO official, who, seeking to protect their domain expertise, expressed skepticism about the model’s value, ultimately stalling progress. 

\noindent \textit{C) Discomfort with Lack of Agency: } Another strategy for navigating trust issues was granting POs greater control early on, such as agency over model outputs and parameters. $P_4$ recalled creating an elaborate software interface at a PO’s request to provide this agency:

\begin{displayquote}
    \textit{Initially, because they didn't know what would happen with the system, they wanted more control. But once they used it and trusted output, they really didn't need the complicated interface. [...] Had we not given them the initial control, they would have not deployed the system [...] so that initial step was probably still useful to get it accepted.} -- $P_4$
\end{displayquote}

In another case, the potential PO is skeptical of the technology and what it promises to deliver itself. $P_6$ recalled an experience working with epidemiologists who were convinced that AI isn't viable in their domain: ``\textit{Some epidemiologists were not even convinced that this can be done [...] so you need to go and convince them that this is a possibility and this will help them do their job better.}'' Overall, trust was defined as a function of time and credibility. Once a project was successfully deployed, interest to collaborate from other 
POs in the same domain is piqued. $P_4$ shared their experience: ``\textit{Once we got one success with wildlife conservation society, then it seemed like it was easier for others to have that trust in us that okay, they are good people, they deliver results and so forth.}''

These reflections, therefore, highlight the nature of building trust in AI4SI research: slow-building, iterative, and earned. Hence, it is vital for AI4SI practitioners to invest in long-term engagement, transparency, and prioritize relation-building just as much as technical development. 

\subsection{Operational Hurdles}\label{sec:operational_hurdles}
Through our analysis, we identified three types of operational hurdles:\\ 
\noindent\textit{4.4.1} $\>$ \textit{Dynamic Environments.}\\
AI4SI projects are developed for POs, which are often characterized by high amounts of variability in both their staff and project priorities. Participants mentioned how quickly the landscape changes in these settings, $P_1$ mentioned: 

\begin{displayquote}
    \textit{The landscape also just keeps changing continuously in this space, [...] so there's a question of how does one keep up with a moving target? Especially when you're doing this in a small academic lab, where the policies on what you're supposed to do and how you're supposed to do it are regularly changing.} -- $P_1$
\end{displayquote}

In rapidly evolving domains  like crisis response, $P_6$ noted tensions between novelty and time constraints: \textit{``because you're dealing with a disaster, you need to be able to respond, and response sometimes doesn't mean having the optimal answer[...] because sometimes, if you find best answers too late, then it's meaningless. So there is a timeliness perspective.''} Interviewees also noted that sudden shifts in policy, stakeholder needs, or organizational structure often made static software ill-suited to evolving deployment contexts, requiring repeated realignment. As $P_3$ emphasized:



\begin{displayquote}
    \textit{The biggest problem is, it [environment] doesn't stay the same. But the software stays the same. So the challenge is that after a while the reason why you don't see wide adoption of a software is because there's lots of variability in deployment space.} -- $P_3$
\end{displayquote}

The dynamic nature wasn't constrained to just the problem environment or organizational structure, participants also mentioned dealing with sudden changes in priorities of their collaborators. $P_6$ shared how their industry PO rapidly shifted focus to generative AI after the launch of ChatGPT, thus deprioritizing ongoing AI4SI efforts.  $P_8$ shared another manifestation of this hurdle by recalling their experience on how political pressure and public scrutiny diverted the attention of their PO while working on a short term project, leaving insufficient time for meaningful progress.



In addition, many nonprofit and public POs face high staff turnover, hindering long-term AI4SI deployment. Reliance on volunteers or short-term staff creates continual retraining needs, often unsupported by project grants. As $P_3$ put it: 
\begin{displayquote}
    \textit{If you build it, can any of their volunteers even use it? [...] You will provide training to some of the people, and it's a very volatile space where people come and go because a lot of them are volunteers. So just because you teach a couple of people doesn't mean it's going to stick.} -- $P_{3}$
\end{displayquote}

Collectively, these examples highlight the uncertainty that characterizes 
in many AI4SI environments. From staff churn at POs to shifting institutional priorities and evolving problem definitions, researchers must adapt their systems midway or risk obsolescence.
Navigating these uncertainties requires not only technical robustness but also the ability to build sustainable long-term partnerships.

\noindent\textit{4.4.2} $\>$ \textit{Data Issues.}\\
Limited access to high-quality data was another recurring challenge, as AI4SI projects often rely on messy, incomplete, or task-misaligned data.Participants noted that some projects never advance because POs lack the necessary data. 
For example, $P_9$ stated \textit{``they're really interested in working on something, but they don't have the data''}. Similarly, $P_{19}$ noted: 


\begin{displayquote}
    \textit{I think, overall half the year to actually figure out what type of data they have. [...] At some point, they were like, ``yeah, we should have this data'', and then [after] multiple months, they realized, they don't have this data. They never collected it, and the whole project was kind of doomed to fail afterwards.} -- $P_{19}$
\end{displayquote}

Throughout interviews, participants emphasized that poor data quality made it challenging to build reliable models that address the PO's needs,
as $P_3$ said \textit{``there were a lot of accuracy challenges as well, primarily, because the data set was quite dirty"}. $P_3$ further noted that POs might not be willing to share their data due to its highly sensitive nature: \textit{``the challenge is that a lot of the data is sensitive, so you can't really share files with someone who's a 3rd party"}.  
This sentiment was echoed by $P_6$, who shared that they spent a year convincing the PO to share data by having students intern there, noting that even when data exists, accessing and organizing it can require significant time and in-person effort:
``\textit{where do we get the data from? It's a long fuse. [...] I had to send a student, and they were not ready to share the data. So the 1st year was very hard. I had to literally send a student for an internship there.}''

In summary, the data collection pipeline is rarely straightforward in AI4SI. Before researchers can begin modeling a solution, they often spend months navigating institutional barriers and data cleaning just to make data analysis possible.

\noindent\textit{4.4.3} $\>$ \textit{Scalability and Maintenance.} \\ 
55\% of AI4SI projects receive less than \$250K in funding \cite{Bankhwal_2024}, so even with promising pilots or field deployments, scaling and long-term maintenance remain major challenges. Since majority of the projects are narrowly scoped in the initial phases, when prompted to extend beyond their original use case, these solutions can turn brittle. As $P_9$ mentioned:


\begin{displayquote}
     \textit{After we get like version 0 of something trying to actually get something that is useful and usable at scale is really hard, [...] the reason these problems are so hard is because they are very locally contextualized. [...] we can't just kind of like multiply and stamp out the exact same version from setting to setting.} -- $P_9$
\end{displayquote}
Even after an intervention has been proven effective, maintaining and supporting it is its own resource-intensive task which requires substantial engineering efforts. $P_4$ reflected on this by stating: \textit{``If you needed a very sophisticated deployment, then you get into serious software engineering. And now you're talking [about] millions of dollars of software engineering contracts, which is obviously not something a university can do or is interested in."}


$P_4$ further shared their experience of successful deployment scenarios, where they mentioned the importance of industry partnerships that are capable of sustaining large-scale software architectures. $P_4$ elaborated:\textit{``If you have a [company partner], collaboration with [that company is] supporting `it' [deployment]. [...] So there's a scale of projects for deployment that can be handled at universities, beyond which we are not equipped to deal with."}

$P_{21}$ expressed concern that sustaining deployed work often falls outside the priorities of academic labs, explaining that, \textit{``deployment may be just enough to write a paper, but beyond that, it also takes a lot of effort to maintain the system, a lot of engineering effort, which is not always valued within the [traditional academic] community''.}
This lack of academic incentive discourages long-term maintenance by research teams. As a result, researchers often focus on demonstrating effectiveness and hope that POs will invest in maintaining the work. As $P_6$ stated \textit{``our interest in many things is to deploy something, show them the work of it, and then obviously, they have to get convinced enough that they maintain it."} Another challenge in sustaining interventions long term is student turnover. $P_1$ shared \textit{``maybe you couldn't find students, or like students who was working on it graduated, and nobody else like kept going"}.

These reflections highlight that scaling and maintaining interventions post-deployment can limit the overall impact of AI4SI projects, as small academic labs often lack the resources, funding, external recognition, and industry partnerships needed to sustain and scale their work. 

\section{Discussion/Possible Solutions}
\noindent \textbf{Contextualizing our Findings with Literature in Allied Fields. }
\citet{aula2023stepping} argue that AI for Good is best understood as part of a broader historical trajectory of computational approaches to social challenges, and several of our findings affirm this view. Within ICT4D specifically, the “publish or perish” tension (Section \ref{sec:4.1.1}) mirrors the observations of \citet{Walsham} and \citet{Harris2016} that institutional incentive structures often discourage the types of practitioner engagement necessary for research to influence policy and practice. Operational challenges related to data access, dynamic environments, and scalability (section \ref{sec:operational_hurdles}) align with \citet{Heeks2008} sustainability, scalability, and evaluation concerns and \citet{Dodson2012} finding that technology-centric approaches often produce unsatisfactory outcomes, complementing the insight of ``divergent priorities'' (Section 4.1.2) which signifies the importance of aligning the research focus and POs needs. Similarly, in the broader T4SG and DSSG space, prior works \cite{pal2017fallacy,Sambasivan2019} highlight how technology-for-good initiatives often fall short of meaningfully engaging with the communities they serve, whether through short-term extractive research engagements that erode community trust (Section 4.3.2) or by treating technological solutions as sufficient to address deeply entrenched social challenges (Section~\ref{sec:operational_hurdles}) \\ 
Furthermore, our findings also extend this literature. 
While prior work in ICT4D and T4SG has documented that technological interventions can misalign with the priorities of the communities they aim to serve \cite{pal2017fallacy,Sambasivan2019,toyama2015geek, brown2019some}, our findings highlight this misalignment can also arise within POs themselves, where leadership's enthusiasm for AI can mask the priorities at the front-line level (Section 4.1.2, $P_4$'s quote). Building on prior ICT4D work on academic incentive structures \cite{Harris2016,Walsham}, we show how the AI/ML publication systems produce a more acute form of these tensions. Although both ICT4D and AI4SI involve interdisciplinary, long-term engagements with their POs, AI4SI work is often evaluated in a fast-paced conference-driven publication system with shorter review cycles than journal based processes, coupled with the expectations for frequent outputs, thus making it challenging to align sustained, deployment-focused work with the standard publication timelines. 

Finally, our findings diverge from prior work in perspective. Much of ICT4D and DSSG work foregrounds the experiences of POs or communities that interact with technological innovation. In contrast, our study primarily focuses on the perspectives of the academic researchers in AI4SI collaborations as our goal is to understand how institutional structures within the research ecosystems shape the development and deployment of these systems. This researcher-centric perspective helps reveal tensions not addressed in allied fields. Though foundational ICT4D work identifies grassroots intermediaries as critical bridge-builders \cite{cecchini2003can}, our findings show that while intermediaries are highly valued, they can simultaneously restrict access for outsiders (Section 4.3.1, $P_{26}$). Likewise, while recent works identify infrastructure and resource constraints as macro-level deployment barriers \cite{Heeks2008,ali2007challenge}, our participants highlighted how the large-scale computational demands of AI/ML systems create a qualitatively different asymmetry where the POs often lack the capacity to independently deploy or host AI tools (Section 4.1.3, $P_{15}$), thus creating sustainability challenges distinct from earlier ICT4D technologies like mobile phones \cite{jackson2012repair} that could be locally maintained by POs. 

\noindent \textbf{Actionable Insights for AI4SI Researchers. }
To help overcome the communication barries and trust issues that prevent deployment (Section \ref{sec:4.2} and \ref{sec:4.3}), participants emphasized the importance of delivering tangible value early through  ``quick wins". This 
was the most stated approach taken by researchers during the initial phases of their AI4SI projects. In this approach, rather than immediately aiming for technical novelty, researchers prioritized delivering small, tangible forms of value to their PO during the initial phase of the project. This could include simple data analysis or running regression models to help answer pressing questions from the PO. These early contributions helped researchers with gaining trust, understanding organizations domain of work, and establishing credibility with their POs.

Beyond delivering early value, standardized documentation frameworks can further help in reducing the communication barriers hindering deployment. 
Tools such as models cards \cite{modelcard} and data cards \cite{datacard} can help reduce miscommunication between researchers and POs. Model cards can help explain model behaviour 
and what its limitations are in a clear and accessible manner. Similarly, data cards can help researchers understand a dataset’s origins, structure, and limitations, fostering shared understanding with POs.

To address the operational challenges of scalability and the structural pressures of academic timelines (Sections \ref{sec:operational_hurdles} and \ref{sec:4.1.1}), some participants recommended leveraging foundation models that provide reusable building blocks across projects, which would help lower the labour demands and enable more scalable deployments, an insight also echoed by \citet{zhao2024foundationmodelbasedmultiagentaccelerateai}. However, others emphasized the importance of having bespoke models to truly capture the intricacies of the problem and create trust with POs.

To navigate the structural challenge that software engineering work is undervalued in academia despite being essential for AI4SI deployment (Sections 4.1.1 and 4.1.4), researchers suggested delegating labor strategically across student levels. Some participants paired PhD students who conducted the core research of the project with master's and undergraduate students to assist with the engineering efforts, such as software development. This delegation of tasks helps make the work more sustainable over time.

Finally, to tackle both the funding constraints and the collaboration hurdles early in partnerships (Sections \ref{sec:4.1.1} and \ref{sec:4.3}), “matching market” style funding opportunities tailored towards the creation of early collaboration opportunities between AI4SI researchers and POs can help. These funding opportunities can provide sufficient incentives to both parties to brainstorm mutually beneficial problems, which can then be submitted as grant proposals. These grants allow researchers and POs to test ideas, align motivations, and build trust without committing resources or funding upfront.
Google India AI for Social Good program \cite{Tambe_Google_2021}, NSF ReDDDot~\cite{NSF_ReDDDOT}, NSF CIVIC \cite{NSF_CIVIC}, and NSF TTP \cite{NSFTTP} are good examples of such programs. The Boring Fund \cite{boringFund} was also noted for supporting the maintenance of projects.

\noindent \textbf{Actionable Insights for Institutions. }To address the structural misalignment between academic incentives and deployment focused work (Section \ref{sec:4.1.1}), participants emphasized expanding the definition of valuable research beyond traditional metrics. 
Most academic systems still value technical novelty and publication count over long-term real-world deployment. Participants noted that valuable research should include both algorithmic novelty and applied work.
 This might include recognizing deployed systems, open-source tools, or public impact as equally important contributions. 

That said, participants acknowledged that the AI4SI landscape is improving. 
Compared to a few years ago, several conferences now offer special tracks for AI4SI projects (e.g., AAMAS \cite{AAMAS}, NeurIPS \cite{NeurIPS2018_AI4SG}, KDD \cite{kdd2023}, IJCAI \cite{ijcai2025} and AAAI \cite{AAAI_2024}) recognizing their significance and deployment challenges, and helping bring greater visibility and legitimacy to this work.
Importantly, there seems to be a welcome shift in how this work is evaluated: expectations for submissions to these tracks are 
evolving to reflect the trade-offs between novelty and real-world relevance highlighted by participants. 
More of this shift is needed to ensure socially grounded work is not sidelined by conventional academic norms.

To tackle the institutional inertia and procedural red-tape that slow AI4SI collaborations (Section 4.1.4), participants suggested implementing 
institutional reforms that would aid researchers in this field. There is a need for stronger support systems, such as university consortia that institutionalize early-stage AI4SI collaboration by connecting partners, managing data use agreements, and aligning goals. 
An example is Carnegie Mellon University's ML for Social Good initiative \cite{CMU_ML4SG}, which provides opportunities for public \& social sector organizations to collaborate with their researchers. These initiatives take the burden of conducting outreach and navigating procedural red tape 
off the shoulders of individual researchers. However, participants also cautioned that consortia may not always succeed. When driven by passionate individuals they can flourish, but without such leadership, they risk becoming ineffective.
 
To address the disciplinary gaps and lack of interdisciplinary training that hinder effective collaboration (Sections 4.1.4, and \ref{sec:4.3}), participants suggested institutions should develop dedicated training programs for students interested in AI4SI. 
Aspiring students often have strong technical foundations but lack training in 
statistical inference, econometrics, and field experiments in social contexts. Participants emphasized the need for better interdisciplinary training involving public policy, qualitative research, or ethics. Some even called for 
dedicated AI4SI programs in CS departments.

\noindent \textbf{Insights for POs. } In order to conduct successful cross-sector collaboration, understanding the contexts in which your partners operate is essential. Rather than offering actionable recommendations as we have for researchers and institutions, this sub-section highlights how POs can benefit from understanding the academic context in which their research partners operate. \citet{hongjin_lin}'s recent work centers the POs perspectives in AI4SI collaborations and provides detailed insights into PO experiences and aspirations, which serves as a valuable complement to this study. Our hope is that understanding the structural constraints AI4SI researchers face (from publish-or-perish to funding limitations) can help POs set realistic expectations and advocate for collaboration structures that work for both parties.






\section{Concluding Thoughts}
AI for Social Impact (AI4SI) represents a growing area of research that seeks to apply AI to real-world problems affecting underserved and marginalized populations. In a world where BigTech continues to dominate the AI landscape by often creating for-profit solutions for first-world problems,
it is imperative that the AI4SI community flourishes to serve as a counterbalance by focusing on the needs of marginalized and underserved populations (who will rarely get attention from BigTech because of the for-profit nature of this industry). The work facilitated by this field may be slower, more complex, and less rewarded, but it remains vital for ensuring that the benefits of AI are more equitably distributed. 

Unfortunately, research in this space is challenging for a variety of reasons, as highlighted in this paper. In general, affecting positive real-world change through research interventions is difficult in any domain, but these challenges are compounded by the structural disincentives in academia.
Research that prioritizes socially relevant deployment over algorithmic novelty is often undervalued by peers, and limited funding further exacerbates the challenges of sustaining long-term, impact-driven work. 
As showcased in this paper, researchers face roadblocks not just during technical development, but also in navigating red tape, collaboration and communication hurdles, aligning incentives with POs, and ensuring their work is sustainable beyond initial pilots. 

The reflections 
in this paper, drawn from the interviews and authors' lived experiences, 
highlight the challenges faced by AI4SI researchers. At the same time, participants’ suggestions and workarounds point to paths for advancing the field. 
By documenting these challenges and lessons, we hope this work raises awareness of the issues AI4SI researchers face and serves as a reference for aspiring practitioners.
Strengthening this field will require collective effort, but this would ensure that the fruits of AI technologies are shared equitably across various communities that need them most.

\section{Limitations}

Due to the non-trivial efforts involved in finding participants for this study, the final sample size is relatively skewed in its demographic configuration. While 15 of the 38 projects in our dataset (Table \ref{sec:appendix_projects}, Appendix \ref{sec:appendix_projects}) were based in the Global South, a majority of the participants were from the Global North, which means our findings may not capture the full range of challenges faced by the researchers worldwide. Although we attempted to recruit participants from multiple continents, scheduling conflicts limited our ability to do so. The incentive structures, collaboration dynamics, and funding constraints facing Global South researchers may differ substantially from those we document.  For example, the publish-or-perish  pressures may manifest differently when top venues are constrained by language barriers and travel costs. Additionally, international collaborations may introduce power asymmetries in which the Global South partner provides data and community access, while the Global North partner retains control over publication and credit. Future work should explicitly recruit Global South researchers to build a more globally inclusive understanding of AI4SI deployment barriers. Along with that, the sampling strategy relied on citation networks, which could bias the results. We acknowledge this limitation and the way it may have shaped the narratives of the paper.  

\section{Ethical Considerations}

We acknowledge the recent critiques of the AI4SI space \cite{green2019good, berendt2018aicommongoodpitfalls, moore2019ai}, specifically those concerned with techno-solutionism: the belief that technological innovations can be used to solve all complex social problems \cite{saetra2023introduction}. While this paper presents the practical challenges to deploying solutions in a real-world context, we do not assume that AI interventions are appropriate or sufficient in all contexts. Structural changes such as implementing informed policies, and sustained investments in public institutions are required to uplift these marginalized and underserved groups. We view AI4SI as one potential tool in this broader ecosystem for change, and emphasize the long-term sustainability and alignment with the community's needs of these systems.  

Finally, in this study, we have ensured to maintain the participants' privacy. The only identifying information that was kept was their emails to send them the demographics survey and updates regarding the study. No personally identifiable information has been disclosed, and all the quotes were anonymized. Any identifying information was removed from the audio transcripts before starting the thematic coding process. To ensure participant's well-being, they were informed that they are allowed to withdraw from the study whenever they pleased to without any consequences. The participants were provided with the goals of the study, and all provided voluntary verbal consent to partake in this study. 

\section{Positionality Statement}
This paper is grounded in the authors' belief that the value of AI4SI work is not defined just by algorithmic novelty, but the successful integration of these systems in real-world contexts that are underserved, resource-constrained, or systemically overlooked. Our normative stance (defined in Section \ref{sec:methods}) prioritizes deployment, sustainability, and alignment with the affected community's needs. We acknowledge that not everyone shares this view, especially in the broader AI community, where the value of a work is often determined via algorithmic novelty and theoretical innovations.  

The authors are researchers working in an academic institution where academic incentives are still largely dependent on publication volume and theoretical novelty. We understand the slow and iterative nature of the work in this field and the tensions that this creates while working in this space. Our normative stance was developed through our lived experiences, trying to deploy our work on the ground. These experiences shape the barriers to deployment presented in this paper. 

By explicitly mentioning our positional stance, we wish to clarify the assumptions that informed our analysis. We do not claim that this is the only way to approach AI4SI, but we believe that the responsible integration of these AI technologies is vital to serve the communities which are often left behind.

\section{Generative AI Disclosure Statement}

We primarily use generative AI to shorten our writing to satisfy the page limit. 

\begin{acks}
Majumdar and Yadav were kindly supported in this research by NSF Grant $\#$2427737.
\end{acks}

\bibliographystyle{ACM-Reference-Format}
\bibliography{sample-base}

\clearpage
\appendix

%


\section{Demographic Details}
\label{sec:appendix_demographics}
The aggregate of the self-reported demographic details of participants are presented in Table \ref{tab:demographic}.

\begin{table}[t]
\centering
\begin{tabular}{|p{4cm}|p{8cm}|}
\hline
\textbf{Demographic Information} & \textbf{Participant Statistics} \\
\hline
Age (years) & Mean: 37.9, Maximum: 60, Minimum: 27 \\
\hline
Gender &  Man (14), Woman (12) \\
\hline
Race & South Asian (9), White (7), East Asian (6), Southeast Asian (2), Black/African American (1), Middle Eastern (1) \\
\hline
Region &  USA (21), India (3), Germany (1), Netherlands (1) \\
\hline
Interview length (minutes) & Mean: 41.36, Maximum: 58, Minimum: 26.5 \\
\hline
Position & Assistant Professor (8), Associate Professor (5),  Senior Research Scientist (2), Research Scientist (2), Professor (2), Researcher/Entrepreneur (2),  Principal Researcher (1), Data Scientist (1),  Program Manager (1), Postdoc (1), PhD Student (1) \\
\hline
Years of Research Experience in AI4SI & Mean: 8.8, Maximum: 18, Minimum: 2 \\
\hline
\end{tabular}
\caption{Participant Demographic Information}
\label{tab:demographic}
\end{table}

\section{AI for Social Impact Projects Discussed during the Interview}
\label{sec:appendix_projects}


The projects mentioned in the interview have been synthesized in Table \ref{tab:AI4SI}. The table groups relevant projects into distinct sectors of AI4SI, such as conservation, public health, safety and security, social justice, food security, and agriculture. There are five distinct deployment statuses: (1) Deployed - the project was successfully deployed; (2) Pilot - test pilots were conducted, but the system wasn't deployed long-term; (3) Used in Advocacy - the results from these studies were used to try to inform policies; (4) Research and Development (R\&D) - preliminary results of the studies were promising and the researchers are in the process of developing a system which may be deployable; (5) Not Deployed - the project could not be deployed. The rows highlighted in blue are the projects that were developed by the authors of this paper.

\begin{longtable}{|p{1.8cm}|p{2cm}|p{4.0cm}|p{1.7cm}|p{1.6cm}|p{1.8cm}|}
\caption*{Overview of selected AI4SI projects and their deployment status, grouped by sector. Rows highlighted in blue indicate projects developed by this paper’s authors.}
\label{tab:AI4SI} \\

\hline
\rowcolor{gray!30}\textbf{Sector} & \textbf{Project Topic} & \textbf{Description} & \textbf{Location} & \textbf{Participant} & \textbf{Deployment Status} \\
\hline
\endfirsthead

\hline
\rowcolor{gray!30}\textbf{Sector} & \textbf{Project Topic} & \textbf{Description} & \textbf{Location} & \textbf{Participant} & \textbf{Deployment Status} \\
\hline
\endhead

\hline
\multicolumn{6}{r}{\textit{Continued on next page}} \\
\hline
\endfoot

\hline
\endlastfoot

\multirow{1}{*}{Conservation}
& AI for Wildlife Security  & Modeling poacher behaviour from historical data and uses game-theoretic planning to optimize patrol strategies. & Africa, Asia & $P_5$, $P_4$, $P_9$, $P_{13}$, $A_4$ & Deployed \\
\cline{2-6}
& AI for Wild Animal Tracking & A computer vision model to identify existing giraffes and track their population trend. & Africa & $P_{20}$ & Deployed \\
\hline

\multirow{2}{*}{Food Security}
& AI for Pallet Planning in Food Banks & Algorithm to estimate pallet needs for food bank orders without exact product dimensions, improving logistical planning under uncertainty. & North America & $P_5$ & Not Deployed \\
\cline{2-6}
& AI for Food Rescue & This project learns from historical data to optimize volunteer notification radii for food rescue via push alerts. & North America & $P_5$ & Deployed \\
\hline

\multirow{5}{*}{Social Justice}
& AI for Homelessness Services & ML and causal inference to improve allocation of limited homelessness services, aiming to reduce reentry and support fair, discretionary decision-making. & North America & $P_1$, $A_4$ & Pilot / Used in Advocacy \\
\cline{2-6}
& AI for Eviction Prevention & Eviction risk models using property, owner, and neighbourhood data to inform proactive outreach strategies that better target at-risk tenants. & North America & $P_1$ & R\&D \\
\cline{2-6}
& AI for Child Trafficking Rescue & Face recognition system to identify and recover trafficked children from brothels. & Asia & $P_3$ & Deployed \\
\cline{2-6}
& AI for Policy Accountability & A collaborative network to improve policy accountability by collecting, organizing, and visualizing data from police departments. & North America & $P_{25}$ & Deployed \\
\cline{2-6}
& \cellcolor{cyan!10} AI for Eviction Forecasting & \cellcolor{cyan!10}Deep learning model to forecast future eviction risk at the census-tract level for better resource allocation. & \cellcolor{cyan!10}North America & \cellcolor{cyan!10}$A_4$ & \cellcolor{cyan!10}Used in Advocacy \\
\hline

\multirow{5}{*}{\shortstack{Safety \&\\Security}}
& AI for Airport Security & Randomized patrol scheduling system to deter threats at LAX airport & North America & $P_4$ & Deployed \\
\cline{2-6}
& AI for Flight Security & Intelligent randomized scheduling for Federal Air Marshals on flights & North America & $P_4$, $A_4$ & Deployed \\
\cline{2-6}
& AI for Maritime Security & Tool for optimizing Coast Guard patrols across ports & North America & $P_4$, $A_4$ & Deployed \\
\cline{2-6}
& AI for Public Transit Safety & Patrol scheduling system used by LA Metro Rail to deter crime & North America & $P_4$ & Deployed \\
\cline{2-6}
& AI for Infrastructure Resilience & Network-based simulation tool for cascading failure analysis in critical infrastructure systems. & North America & $P_6$ & Deployed \\
\hline

\multirow{1}{*}{Agriculture}
&\cellcolor{cyan!10} AI for Climate-Resilient Agriculture  & \cellcolor{cyan!10} Spatiotemporal forecasting system to help East African farmers manage locust outbreaks and irrigation needs using deep learning and remote sensing. & \cellcolor{cyan!10}Africa & \cellcolor{cyan!10}$A_4$ & \cellcolor{cyan!10} Deployed \\
\cline{2-6}
& AI for Agricultural Remote Sensing & ML models and Datasets for crop type classification, phenology prediction, and yield forecasting for framers coping with aridification. & Asia & $P_{22}$, $A_4$ & Not Deployed \\
\hline

\multirow{7}{*}{Public Health}
& AI for Pandemic Forecasting & Deep learning system for real-time COVID-19 forecasts using noisy, heterogeneous signals. & North America & $P_6$ & Deployed \\
\cline{2-6}
& \cellcolor{cyan!10} AI for Maternal Health & \cellcolor{cyan!10} NLP-based SMS triage system to prioritize maternal care for pregnant women. & \cellcolor{cyan!10} Africa & \cellcolor{cyan!10}$A_4, A_2$ & \cellcolor{cyan!10}Deployed \\
\cline{2-6}
& AI for HIV Prevention & Social network–optimized peer leader selection to reduce HIV risk behaviors among homeless youth. & North America & $P_2$ & Deployed \\
\cline{2-6}
& \cellcolor{cyan!10} AI for COVID-19 Testing Strategy & \cellcolor{cyan!10} Adaptive POMDP-based model to optimize testing kit allocation for symptomatic and asymptomatic individuals. & \cellcolor{cyan!10} North America & \cellcolor{cyan!10} $A_4$ & \cellcolor{cyan!10} Not Deployed \\
\cline{2-6}
& AI for Infection Control & Hybrid neural and agent-based model to identify and forecast hospital-acquired infections. & North America & $P_6$ & R\&D \\
\cline{2-6}
&\cellcolor{cyan!10} AI for Substance Use Prevention &\cellcolor{cyan!10} ML tools that identify substance use risk among homeless youth using social media and survey data. &\cellcolor{cyan!10} North America &\cellcolor{cyan!10} $A_4$ &\cellcolor{cyan!10}  Not Deployed \\
\cline{2-6}
& AI for Fair Pain Assessment & AI model using X-rays to better capture pain severity in underserved groups. & North America & $P_{11}$ & Not Deployed \\
\cline{2-6}
& AI for Mobile Clinic Allocation & AI model to help make predictions of the mobile clinics location. & North America & $P_{16}$ & Not Deployed \\
\cline{2-6}
& AI for Vaccine Estimation & AI model to predict child vaccine demand for local clinics in Kenya. & Africa & $P_{16}$ & Deployed \\

\cline{2-6}
& AI for Medication Adherence & System to monitor and improve medication adherence for tuberculosis patients. & Asia & $P_{24}$ & Deployed \\
\cline{2-6}
& AI for COVID-19 Spread Prediction & A decision-support tool for policymakers that uses epidemiological models to quantify and forecast the impact of mobility changes on COVID-19 infection rates. & North America & $P_{26}$ & Not Deployed \\
\cline{2-6}
& AI for Improving Maternal Health Outreach & A system for allocating limited resources, such as health worker time, to achieve better health outcomes (in a maternal and child health). & Asia & $P_{15}, P_4$ & Deployed \\
\cline{2-6}
& AI for Respiratory Disease Diagnosis & Deep learning algorithms to help with diagnosing the chest movements to understand what kinds of respiratory diseases they belong to. & North America & $P_{10}$ & Not Deployed \\
\cline{2-6}
& AI for Pregnancy and Infancy Issues & A intervention allocation algorithm to improve pregnancy and infancy issues in India. & Asia & $P_{19}$ & Deployed \\
\cline{2-6}
& AI for Health Intervention & AI tool to optimize health interventions and increase child vaccination uptake. & Africa & $P_{23}$ & Deployed \\
\hline
\multirow{2}{*}{\shortstack{Environment}}
& AI for Noncompliance Detection & ML algorithm to help detect noncompliance with federal law related to water pollution. & Asia & $P_7$ & Deployed \\
\cline{2-6}
& AI for Affordable Solar Energy. & ML system to support affordable solar energy by predicting repayment patterns for users. & Asia & $P_8$ & Not Deployed \\
\hline
\multirow{2}{*}{\shortstack{Linguistic\\Diversity}}
& AI for Equity in Linguistic Diversity & Bias detection model to foster minority opinion among low-income communities. & Asia & $P_{14}$ & Not Deployed \\
\cline{2-6}
& AI for Misgendering Reduction in LLMs & An algorithm to reduce gender prediction errors based on different languages and cultures. & Asia & $P_{21}$ & Deployed \\
\hline
\multirow{1}{*}{\shortstack{Education}}
& AI for Equitable Education & Weakly supervised deep learning techniques with satellite images to improve school connectivity. & Africa & $P_{18}$ & Deployed \\
\hline
\multirow{5}{*}{\shortstack{Other Social\\Issues}}
& AI for Humanitarian Informing & The ML model used for text summarization on documents related to the global crises and disasters. & Global & $P_{12}$ & Deployed \\
\cline{2-6}
& AI for Public Transportation & A public transportation system that utilizes AI agents in conjunction with platform constraints to provide better public service for users. & North America & $P_{17}$ & R\&D \\
\cline{2-6}

\end{longtable}

\noindent Descriptions of each project are provided in the following subsections: 

\subsection{Conservation}
\textbf{AI for Wildlife Security.}
This project improves anti-poaching efforts by guiding patrols through high-risk wildlife areas using terrain, animal activity, and human behavior data. It generates strategic, realistic routes that adapt over time, balancing exploration with known hotspots. Pilot tests were conducted in Africa, and the system was deployed, it helps rangers cover critical zones efficiently and deter poaching \cite{nguyen2015making,gholami2019don}.
\\
\\
\noindent\textbf{AI for Wild Animal Tracking}. This project focuses on the identification and re-identification of Maasai giraffes in a country where their population had fallen by 52\% in recent decades. Using computer vision, the team developed a scalable image-matching algorithm to help ecologists catalog individual giraffes from thousands of field photos captured over the years. 

\subsection{Food Security} 
\noindent \textbf{AI for Pallet Planning in
Food Banks. } This project
concentrates on estimating the resources required to complete food bank orders. It proposes an algorithm to estimate the number of pallets needed for an order based on the quantity of products, the known products-per-tier, and tiers-per-pallet values. \\

\noindent \textbf{AI for Food Rescue.} This line of work addresses volunteer coordination challenges in food rescue operations through successive AI-based interventions. The first project developed a predictive model to assess whether a food rescue would be claimed, helping dispatchers plan interventions more efficiently and reduce unnecessary notifications. Building on this, the second project created a recommender system that selects and diversifies which volunteers receive notifications, significantly improving task claim rates and reducing volunteer fatigue. The most recent project goes a step further by predicting the difficulty of rescue tasks using both structured and text data, offering interpretable insights to better match tasks with volunteers’ experience levels and support long-term engagement.

\subsection{Social Justice}
\noindent \textbf{AI for Homelessness Services.}
AI has been used to improve how limited homelessness resources are allocated. Early work showed that predicting individual outcomes for different support services can reduce repeat homelessness cases and improve overall system efficiency. Later studies found that showing these predictions to decision-makers can shift how they prioritize cases, raising important questions about fairness and institutional goals. Most recently, machine learning has been used to analyze caseworker discretion, revealing that while many decisions follow consistent patterns, discretionary choices often improve outcomes for less obvious cases, highlighting the value of expert judgment alongside algorithms \cite{wilder2021clinical,yadav2015preventing,yadav2016psinet,yadav2016using,yadav2016pomdps,soriano2016simultaneous,yadav2016healer,yadav2017using,wilder2017uncharted,yadav2017maximizing,yadav2018influence,hu2018activating,yadav2018bridging,rice2018piloting,yadav2018please,rahmattalabi2019exploring,yadav2021optimal,wilder2021clinical,tabar2020identifying,dou2021harnessing,rice2021peer}.\\

\noindent \textbf{AI for Eviction Prevention.}
This project addresses the question of how useful the eviction predictions are in targeting outreach efforts. They create risk scores by completing an eviction prediction assignment, and then utilize these risk ratings to develop targeted outreach strategies. Experiment results demonstrate that the risk scores are effective, allowing a hypothetical caseworker team to reach more properties that are at risk of eviction.\\

\noindent \textbf{AI for Child Trafficking Rescue.} This project
prioritizes the child trafficking issue and introduces two
systems to retrieve trafficked children from brothels in India using computer vision. They specify the model's performance is contrasted with a number of cutting-edge techniques.\\

\noindent \textbf{AI for Eviction Forecasting.} \cite{Maryam2,evictiontabar2022_example} focuses on the prevention of tenant eviction and develops a multi-view deep neural network model to forecast the number of tenants at risk of getting formally evicted (at the census tract level) n months into the future.\\ 

\noindent \textbf{AI for Policy
Accountability.} This project focuses on gaining a deeper understanding of the legal landscape of police accountability in a specific county. It establishes an open, collaborative network designed to address challenges such as data inaccessibility, data complexity, and the fragmentation of data across different municipalities, ultimately making this previously inaccessible information available to the public.

\subsection{Safety \& Security}
\noindent \textbf{AI for Airport Security.}
This project discusses the Los Angeles airport security protection problem and applies a game-theoretic model for this patrolling/monitoring problem. The research question was focused on where and when you set up checkpoints, and do canine patrols and do so in a principled fashion, knowing that the adversaries conduct surveillance; and was solved using the concept of randomization.\\ 

\noindent \textbf{AI for Flight Security.}
This project focuses on security issues at airport terminals and air marshals on international flights. They propose two applications to solve these issues accordingly. These applications are also deployed in different real-world scenarios. This study looks at the information, design decisions, and assessment standards that are essential to creating these applications \cite{nguyen2019tackling,nguyen2020decoding}. \\


\noindent \textbf{AI for Maritime Security.}
This project focuses on the security allocation problem for the United States Coast Guard and provides a game-theoretic system, as the solution. They provide findings from an actual experiment that validates the suggested model and, more significantly, quantifies the advantages of security games for allocating security resources \cite{shieh2015extended}.\\

\noindent \textbf{AI for Public Transit Safety.}
This project focuses on a real transit system and introduces a new general security resource allocation in dynamic, unpredictable settings using the Bayesian Stackelberg game model. By incorporating a Markov Decision Process, it generates adaptive schedules that handle real-world disruptions. Field deployment on the LA Metro showed improved fare evasion outcomes over static methods.\\

\noindent \textbf{AI for Infrastructure Resilience.}
This project focuses on the Critical Infrastructure Systems (CIS) protection problem. They creates the CIS analysis tool, which employs heterogeneous network analysis to let users: (a) see the country's vital infrastructures and their intricate interdependencies; (b) find network vulnerabilities; and (c) run quick simulations and other analyses to find answers to "what-if" questions.

\subsection{Agriculture}
\noindent \textbf{AI for Climate-Resilient Agriculture.} ~\cite{tabar2022mitigating} creates forecasting tools for three commonly used crop-productivity-related variables (i.e., actual evapotranspiration, reference evapotranspiration, and net primary production) so that farmers can utilize them to make necessary adjustments on their farms. Furthermore, ~\cite{tabar2021plan} develops a machine learning method to predict locust migratory patterns across East Africa at high spatial and temporal resolution to help tackle the locust swarming crisis. \\ 
\\
\noindent\textbf{AI for Agricultural Remote Sensing}. This project supports smallholder farmers in India facing aridification due to climate change by using AI and remote sensing to track crop types, growth cycles, and yields. The system enables large-scale monitoring of agricultural trends, helping inform sustainable farming practices and food security policies \cite{guo2020improving,gomes2019computational}.

\subsection{Public Health}
\noindent \textbf{AI for Pandemic Forecasting.}
This project
addresses the real-time emerging pandemic forcasting problem and presents an operational deep learning framework targeted for real-time COVID-19 forecasting.\\

\noindent \textbf{AI for Maternal Health.} 
~\cite{10.1609/aaai.v37i12.26709} focuses on the maternal health issues in Kenya and creates an NLP framework to handle code-mixed texts~\cite{zhang2025chaillmsimprovingcodemixed,wang2025monolingualmultilingualmisinformationdetection} and provide timely maternal and Neonatal care (through the emergency level prediction task).\\

\noindent \textbf{AI for HIV Prevention.}
This project emphasizes the intervention for HIV prevention among homeless youth. They answer the question of which young people ought to be chosen as peer leaders to optimize the intervention's overall effect. To maximize these social network interventions in a community health context, they create an artificial intelligence system.\\

\noindent \textbf{AI for COVID-19 Testing Strategy.} 
~\cite{liang2020let} points out robust testing program is necessary for containing the spread
of COVID-19 infections. The project defines the optimal testing problem, presents the DOCTOR POMDP model, and uses a scalable technique based on Monte Carlo tree search to solve the DOCTOR POMDP. \\

\noindent \textbf{AI for Infection Control.}
This project focuses on the problem of Healthcare-associated infections from multi-drug resistant organisms . They proposes a novel framework that integrates neural networks and agent-based models to leverage the strengths of both mathematical modeling and machine learning methods.\\

\noindent \textbf{AI for Substance Use Prevention.} 
~\cite{tabar2020identifying} focuses on substance use disorder (SUD) where the project develops precise machine learning models to forecast homeless youths' vulnerability to SUD by analyzing environmental, psychological, and other aspects linked to SUD in this population. In conjunction, ~\cite{dou2021harnessing} demonstrates how to detect substance use among homeless youth based on texts from social media.\\

\noindent\textbf{AI for Fair Pain Assessment}. This project investigates the disparities in chronic pain assessment among underserved populations. Using deep learning on medical images, the team developed an algorithm that better captures the severity of pain than traditional clinical assessments, narrowing unexplained gaps in reported pain across racial and socioeconomic groups. \\

\noindent\textbf{AI for Mobile Clinic Allocation}. This project focuses on improving the operations of mobile health clinics serving underserved communities. Using publicly available data, such as weather and transportation patterns with clinic records, the team developed predictive models to forecast daily demand at different clinic locations to help with resource allocation and scheduling of staff to provide more accessible care. \\

\noindent\textbf{AI for Vaccine Estimation}. This project focuses on the fluctuations in child vaccinations in underserved regions. By leveraging ML methods the team was able to develop a model to predict fluctuations in vaccine demand affected by factors such as climate, mobility, and seasonal patterns to help clinics better plan outreach and resource allocation. \\

\noindent\textbf{AI for Medication Adherence}. This project developed a digital medicine adherence system for patients affected by tuberculosis. The system monitors the daily doseage of the treatment, and send timely follow-ups, with the goal of reducing drug resistance and improving cure rates. \\

\noindent\textbf{AI for COVID-19 Spread Prediction}. This project developed an AI-driven tool to model the effects of mobility restrictions during the pandemic. The goal was to provide policymakers with a tool to help evaluate reopening strategies and manage infection spread. \\

\noindent\textbf{AI for Improving Maternal Health Outreach}. This project focuses on strengthening the maternal and child health program. It works to improve patient engagement by optimizing outreach calls with decision-focused learning, and thus aiming to support equitable access to preventive care for underserved communities.\\

\noindent\textbf{AI for Respiratory Disease Diagnosis}. This project leveraged deep learning methods to analyze infant chest movements for early detection of respiratory diseases. Alongside technical work, the project also focused on how front-line workers perceive AI-enabled health tools, highlighting the importance of explainability and usability for real-world adoption. \\

\noindent \textbf{AI for Pregnancy and Infancy Issues.} This project prioritizes the listenership constructions to effectively improve the delivery of maternal and child health information to pregnant women. They utilize AI to identify the most beneficial listenership from an intervention.\\ 

\noindent \textbf{AI for Health Intervention} This project concentrates on vaccination uptake problem, especially for newborns and young children. the authors propose  a new framework (through linear programming) to optimize the vaccination intervention.\\

\subsection{Environment}

\noindent \textbf{AI for Noncompliance Detection.} This project focuses on the noncompliance detection problem with federal law (in particular for the water pollution domain). They explore allocation schemes between different machine learning models, demonstrating that algorithmic design can clarify the sources of administrative policy discretion.\\

\noindent \textbf{AI Affordable Solar Energy.} This project applies AI to expand access to clean solar energy in underserved communities. By developing predictive models of customer repayment systems, it helps providers offer affordable financing while ensuring sustainability of solar programs.

\subsection{Linguistic Diversity}

\noindent \textbf{AI for Equity in Linguistic Diversity.} This project conducts experiments to identify gender-biased Hindi sentences. It develops a gender bias detection model and provides recommendations for mitigating such bias, with a particular focus on promoting community rights and inclusion.\\

\noindent \textbf{AI for Misgendering Reduction among LLM Applications.} This project addresses misgendering issues in texts written in non-majority languages. It develops algorithms to assess and mitigate misgendering across more than 40 languages and dialects using a participatory design approach.

\subsection{Education}

\noindent \textbf{AI for Equitable
Education.} This project focuses on improving global school connectivity to promote inclusive and equitable quality education. It proposes a weakly supervised deep learning approach for identifying school locations in high-resolution satellite images.

\subsection{Other Social Issues}
\noindent \textbf{AI for Humanitarian Informing.} This project focuses on efficiently collecting humanitarian information from around the world. It designs an AI-powered issue detection service to flag sensitive or inappropriate content, along with AI-driven summarization tools that transform lengthy reports into concise overviews to support faster and more informed decision-making.\\

\noindent \textbf{AI for Public Transportation.} This project builds AI-based scheduling tools to improve the reliability of on-demand transportation services for communities with limited access to reliable transportation by accounting for real-world operational constraints to support more consistent service delivery. 


\end{document}